\documentclass[11pt]{article}
\usepackage{amsfonts,pdfsync}
\usepackage{amsthm}
\usepackage{authblk}
\usepackage{amsmath}
\usepackage{graphicx}
\usepackage{comment}
\usepackage[usenames,dvipsnames]{xcolor}
\usepackage{enumerate}
\usepackage{extarrows}
\usepackage{hyperref}
\usepackage[round, authoryear]{natbib}
\usepackage{stackrel}
\usepackage{centernot}
\usepackage{caption}
\usepackage{subcaption}
\usepackage{fancyhdr}
\newcommand{\bfM}{{\bf M}}
\newcommand{\bfm}{{\bf m}}

\let\quoteOLD\quote
\def\quote{\quoteOLD\small}

\definecolor{labelkey}{cmyk}{0,0.8,1,0.5}
\definecolor{refkey}{cmyk}{0,0.8,1,0.5}

\newtheorem{theorem}{Theorem}

\newtheorem{definition}{Definition}
\newtheorem{lemma}{Lemma}

\numberwithin{equation}{section}
\numberwithin{theorem}{section}
\numberwithin{corollary}{section}
\numberwithin{proposition}{section}
\numberwithin{lemma}{section}
\numberwithin{definition}{section}
\numberwithin{remark}{section}

\newtheorem{example0}{\sc Example}[subsection]

\newcommand{\SE}{\rm SE}

\makeatletter
\def\th@newremark{\th@remark\thm@headfont{\bfseries}}
\makeatletter

\def\nexto{\kern -0.54em}

\newcommand{\capfont}{\setlength{\baselineskip}{0.8\baselineskip}\small}

\def\boxit#1{\vbox{\hrule\hbox{\vrule\kern6pt
          \vbox{\kern6pt#1\kern6pt}\kern6pt\vrule}\hrule}}

\newcommand{\JJJ}{{\cal J}}

\newcommand{\BB}{\mathbb{B}}

\newcommand{\NN}{\mathbb{N}}

\newcommand{\R}{\Bbb{R}}

\newcommand{\N}{\Bbb{N}}

\newcommand{\wt}{\tilde}
\newcommand{\wh}{\widehat}
\newcommand{\what}{\widehat}

\newcommand{\be}{\begin{equation}}\newcommand{\ee}{\end{equation}}
\newcommand{\bea}{\begin{eqnarray}}\newcommand{\eea}{\end{eqnarray}}
\newcommand{\bean}{\begin{eqnarray*}}\newcommand{\eean}{\end{eqnarray*}}
\newcommand{\ben}{\begin{equation*}}\newcommand{\een}{\end{equation*}}
\newcommand{\ba}{\begin{aligned}}\newcommand{\ea}{\end{aligned}}

\newcommand{\halmos}{\quad\hfill\mbox{$\Box$}}

\newcommand{\PD}{\textbf{\rm PD}}

\newcommand{\rmd}{{\rm d}}

\newcommand{\PP}{\mathbb{P}}
\newcommand{\EE}{\mathbb{E}}

\begin{document}

\title{Models for Genetic Diversity Generated by Negative Binomial Point Processes}
\vspace{-1ex}
\author[1]{Yuguang F. Ipsen\vspace{-2ex}\thanks{ Email: Yuguang.Ipsen@anu.edu.au}}
\author[2]{Soudabeh Shemehsavar\thanks{Email: Shemehsavar@khayam.ut.ac.ir.
 Dr Shemehsavar thanks the Research School of Finance,  Actuarial Studies \&  Statistics for facilities and support during a sabbatical visit there. 
}}
\author[1]{Ross A. Maller\thanks{Research partially supported by ARC Grant DP1092502; Corresponding author: Email: Ross.Maller@anu.edu.au}\vspace{-1ex}}
\affil[1]{Research School of Finance,  Actuarial Studies \&  Statistics,
Australian National University}
\affil[2]{School of Mathematics, Statistics \& Computer Sciences,
 University of Tehran}

\date{\vspace{-3ex}\today}
\vspace{-8ex}
\maketitle

\vspace{-6ex}

\begin{abstract}
We develop a  model  based on a generalised  Poisson-Dirichlet distribution for the analysis of genetic diversity, %% and  species abundance data.  % for the new model. To illustrate its use, %Two data sets  are analysed. %%One contains 
 and illustrate its use on 
microsatellite data for the genus {\it Dasyurus}  (the quoll,  a marsupial carnivore listed as near-threatened in Australia).
%;  the second is some species abundance data % consisting of  observations  on 
%for certain species of  tropical beetles. 
Our class  of distributions, termed $\PD_\alpha^{(r)}$,  is constructed from  a negative binomial point process,  generalizing  the usual  one-parameter  $\PD_\alpha$ model,
% an existing version of a Poisson-Dirichlet distribution, 
which is constructed from  a Poisson point process. 
Both  models have at their heart a Stable$(\alpha)$ process, but in $\PD_\alpha^{(r)}$, an extra  parameter $r>0$ adds flexibility, analogous to the way  the negative binomial distribution allows for ``overdispersion" %beyond that of the Poisson.    
 in the analysis of count data. 
A key result obtained is a generalised version of Ewens' sampling formula for  $\PD_\alpha^{(r)}$.

We outline the theoretical basis for the model, and, for the quolls data, estimate the  parameters $\alpha$ and $r$ by least squares, showing how the extra parameter $r$ aids in the interpretability  of the data
 by comparison with the standard $\PD_\alpha$ model.
 The methods potentially have implications for the management and conservation of threatened populations.

\end{abstract}

\noindent {\small {\bf Keywords:}}
 Poisson-Dirichlet distribution, negative binomial point process, Ewens sampling formula,
 %Bayesian models.  %trimmed $\alpha$-stable subordinator,
genetic diversity of quolls
%, species abundance models, 
%random discrete distributions, exchangeable partition probability function,  

%\noindent{\small {\bf 2010 Mathematics Subject Classification:}  Primary  60G51, 60G52, 60G55.}

\section{Introduction}\label{Intro}
In this era of big data analyses there still remain situations where sample sizes are necessarily small and numbers of variables are limited.
This is especially the case in conservation genetics where study populations are typically restricted and
data is difficult and expensive to obtain. 
The level of genetic variation among members of a population and how it differs between populations %of similar species 
are fundamental objects of interest and of special significance for guiding conservation  strategies, and it is important to make optimal use of what observations are available. 
Consequently there is a need for specialized techniques for the analysis and interpretation of gene sampling data in conservation genetics studies. 

In the past a number of models for genetic diversity have been based on the Poisson-Dirichlet distribution and the  associated Ewens sampling formula (\cite{Ewens1972}).
As its name suggests, the Poisson-Dirichlet has as an underlying building block 
a Poisson point process. 
A two-parameter class of distributions,  recently proposed  by  \cite{trimfrenzy, IpsenMaller2018}, %and \cite{IpsenMaller2018}, 
is based on  a negative binomial rather than a Poisson point process, an extension which   adds flexibility for modelling purposes, analogous to the way  the negative binomial allows for over-dispersion  in the analysis of count data.

%Applications of these kinds of models are not restricted to genetic diversity studies. A closely related %and vigorous 
%stream of research concentrates more generally on biodiversity and is concerned with species abundance distributions.
%We consider such applications also, to illustrate the range of possibilities of the model. %% using some other relevant types of data sets.

In this paper we use genetic data on species of Australian quolls 
%as well as some species abundance data 
to illustrate how the modified model can be used to quantify diversity within a population and make soundly based statistical comparisons between populations.
Our analysis of the small quolls data set detects differences not found with standard types of analysis.
Apart from the data analyses, we present theoretical rationales for the proposed methodology and  study the properties of the parameter estimates by simulations.  

%The remainder of the paper is arranged as follows.
%In Subsection \ref{pkalpha} the derivation of $\PD_\alpha^{(r)}$ from a trimmed Stable$(\alpha)$ subordinator 
%and its connection with a negative binomial point process are outlined. 
%The random partitions generated by the $\PK^{(r)}$ class are studied in Section \ref{sect:EPPF} by presenting  in Theorem \ref{EPPF1} two formulae for the EPPF. 
%Results from Theorem \ref{EPPF1} are specialised to the subclass $\PD_\alpha^{(r)}$ to derive the generalised Ewens sampling formula in Section \ref{ss3}, which is then used for model fitting to  data examples in Section \ref{data}. Section \ref{sims} reports on investigations of the properties of the estimates through simulations, and  Section \ref{Dis} concludes with a general discussion. Proofs and some verifications are relegated to an Appendix.

\subsection{Quolls Data}\label{dcs}
The quoll is a marsupial carnivore which survives in a limited number of locations across the Australian continent,  New Guinea and Tasmania. In Australia all species are listed as near-threatened. 
A recent paper by \cite{Firestone2000} contains genetic data on 4 species of quoll, a subset of 
which we analyse for genetic diversity. 
% distributed in 20 locations. % across mainland Australia and Tasmania.
Due to its  near-threatened status the numbers of animals sampled are necessarily small and the genetics data is accordingly  restricted.
We concentrate on samples of a single species (the western quoll) from 2 locations in Western Australia and analyse microsatellite sample data on 2 loci. There were a maximum of 15  and 34 animals available in the 2 geographical locations.

%We also present the results of an analysis of a larger  data set consisting of  observations made by Janzen (1973) on tropical insects (species of beetles)   in 25 different sites in Costa Rica and the Caribbean Islands. This larger set
%(560 and 737  beetles in 2 groups) allows a more extensive investigation of features of the model which among other things suggests a natural interpretation of the parameters. 

\section{Methods}\label{meth}\
Assuming ``selective neutrality" in a gene fitness model,  \cite{Ewens1972}  derived a distribution for  vectors of the type  $\bfM=(M_1,M_2,\ldots, M_n)$, where $M_j$ counts the number of times an allele of type $j$ occurs in a sample of size $n$.
This kind of model is appropriate for experimental data in which the number of types, or what types exist, in a population, are unknown, but it is possible in a sample of size $n$, say,  to recognize that there are $k$ types with a number $M_j$ of type $j$, $j=1,2,\ldots, n$. 
(See  \cite{ watt1974} for further discussion.) 
In   \cite{Fisher1943},
R.A. Fisher, also strongly motivated by genetics,   was concerned  with measuring species diversity using a model based on  the negative binomial distribution. 
Fundamentally, the %finite dimensional 
Dirichlet distribution   %of vectors on the simplex  %%, $\nabla_n$,  
underlies both approaches. 

%As a second example,
These two important streams of statistical modelling methodology came together in the paper of  \cite{kingman1975}.
By taking a limit of  finite Dirichlet distributions, analogous to the way in which the Poisson distribution is obtained as a limit of binomials,  Kingman derived a class of models referred to as the Poisson-Dirichlet distributions. One member of these is %one  termed 
the  $\PD_\alpha$ distribution, which is obtained as the limit of the ranked vector of normalised jumps of a driftless stable subordinator with a single parameter $\alpha$  in $(0,1)$). 
 The subordinator jumps can be regarded as a Poisson point process in time and space, and taking this point of view opens up a wide variety of applicable theory which has subsequently been expanded in many directions. 
 %approach to derived a two-parameter class, denoted by $\PD(\alpha,\theta)$,
% which incorporates $\PD_\alpha=\PD(\alpha,0)$ and $\PD(\theta)=\PD(0, \theta)$ as subclasses. See also \cite{perman_thesis} and  
In particular, this approach was further developed  for the $\PD_\alpha$  model by \cite{perman_thesis,  perman1993}.
%(see also \cite{perman_thesis}).
%%The $\PD_\alpha$ model has been applied in a variety of contexts (refs??).

Recently \cite{trimfrenzy, IpsenMaller2018} extended the Kingman-Perman  $\PD_\alpha$ model by replacing the  Poisson point process with a   negative binomial  point process, thereby bringing in an extra parameter, $r>0$.
The new class, termed  $\PD_\alpha^{(r)}$, 
is  amenable to analysis and among other results a generalised Ewens sampling formulae can be obtained for it.
 In what follows we outline the basic structure underlying   $\PD_\alpha^{(r)}$, 
 especially relating to  the negative binomial point process, suggest a  weighted least squares method of fitting it to data, and apply the method to analyse the quolls data. The resulting analyses provide extra insight over and above that obtainable from fitting a 
  $\PD_\alpha$ model. 
 Some simulations verify that the estimators have reasonable properties and this is further demonstrated  in  a theoretical appendix.

\subsection{Data Structure and   $\PD_\alpha^{(r)}$ Model}\label{ss3}
%%The Ewens sampling formulae for a $\PD^{(r)}_\alpha$-partition is needed for the data analysis in Section \ref{data}.
Consider a sample of size $n$ from a population with, nominally,  infinitely many  blocks, whose proportions are governed by a  distribution on the simplex  $\nabla_{\infty}:=\{x_i\ge 0, i=1,2,\ldots, \sum_{i\ge 1}x_i=1\}$.
In  the quolls data,  ``blocks" represent allele types.
%% in the quolls data. %%, or species, in the insects data.

Corresponding to a given partition 
%$\Pi_n=(A_1,\ldots,A_k)$ 
of $\N_n:=\{1,2,\ldots,n\}$ into $k$ blocks, let  $M_j$ denote the number of blocks %(species)of $\Pi_n$
in the sample which have exactly $j$ representatives.  
The vector
$\bfM = (M_1,\ldots,M_n)$
consists of  non-negative, integer-valued random variables satisfying 
 $\sum_{j=1}^{n}j M_j = n$ (sample size) and  $\sum_{j=1}^{n}M_j=k$ (the number of distinct alleles in the sample). %%is the {\it blocks count vector}.  
Let $n_i$ be the number of representatives (alleles) of block type $i$ (allele type $i$), $1\le i\le k$, in the sample, with $\sum_{i=1}^k n_i=n$.
%Given  a vector $\bfm=(m_1,m_2,\ldots,m_n)$
%of  non-negative integers with $\sum_{j=1}^{n} m_j=k$ and 
%$\sum_{j=1}^{n}j m_j=n$,  let

 Ewens' sampling formula (\cite{Ewens1972}) %%%(\cite{Antoniak:1969},)(\cite{Ewens1972})
gives the probability of a particular realization of the blocks count vector $\bfM$
under the assumption of selective neutrality.
% sampled from the hypothesised  distribution.
%We can specialise Theorem \ref{EPPF1} the EPPF for  $\PD_\alpha^{(r)}$ and use it obtain the Ewen's sampling formulae for $\PD_\alpha^{(r)}$. 
Our aim is to present a generalised version of it for $\PD_\alpha^{(r)}$
%, in a form convenient for computations, 
and illustrate its usefulness for data analysis.

\subsection{Negative Binomial Point Processes }\label{nbpp}
The $\PD_\alpha^{(r)}$ class is based on a negative binomial point process (NBPP), denoted by $\BB^{(r)}$, as defined by \cite{Gregoire1984}.
This point process has Laplace functional
\be\label{1.1}
\Phi(f) = \EE\Big(e^{-\BB^{(r)}(f)} \Big) =\Big(1 + \int_{\R_+}(1-e^{-f(x)})\Lambda(\rmd x) \Big)^{-r},
\ee
for  nonnegative measurable functions $f$ on $\R_+$.
In \eqref{1.1}, $r>0$ is a parameter and  $\Lambda$ is a measure on the Borel sets of $\R_+$. %%, locally finite at infinity.
%Denote the distribution of such a $\BB^{(r)}$  by $\BN(r, \Lambda)$ or by $\BN(r, \rho)$ if $\Lambda$ admits a density $\rho$.
%See \cite{Gregoire1984} for the definition and basic properties of an NBPP.
Gregoire devised the NBPP in order to %show that it is possible to
 define a process for which all finite-dimensional distributions associated with disjoint bounded Borel sets
in $\R_+$  are negative binomial.

\subsection{The $\PD_\alpha^{(r)}$ Class}\label{pkalpha}
%One notable subclass of  $\PK^{(r)}$ is  $\PD_\alpha^{(r)}$, which arises naturally from an $r$-trimmed $\alpha$-stable subordinator.
%% and is not in the two-parameter Poisson-Dirichlet $\PD(\alpha, \theta)$ class of distributions.  We discuss its derivation next.
%In this section we show how $\PD_\alpha^{(r)}$ can be constructed using {\it ratios}
%of jumps of a trimmed stable subordinator, rather than the jumps themselves. 
%The  negative binomial point process then  arises in a natural way.
%% from this analysis. See also Section 5 of \cite{IpsenMaller2018}.

To construct  $\PD_\alpha^{(r)}$, we choose a measure having density 
\be\label{LM2}
\rho(x)\rmd x:= 
\alpha x^{-\alpha-1}\rmd x, \quad \text{for some}\  0<\alpha<1,
\ee
%and Laplace transform satisfying
%\ben %%\label{psdef}
%-\log \EE e^{-\lambda S_t}   %= e^{-t\Psi(\lambda)}, \quad \text{where} \quad\Psi(\lambda)
% = \int_0^\infty \big(1-e^{-\lambda x}\big)  \Lambda(\rmd x),\
%\lambda>0.
%\een
and take a driftless stable subordinator $(S_t)_{t\geq{0}}$ of index $\alpha$  with $S_0=0$,   whose  L\'evy measure is given by \eqref{LM2}. 
Its  jump process is
$(\Delta S_t := S_t-S_{t-})_{t>0}$, with $\Delta S_0=0$,
and the jumps up till time 1 are ordered as $\Delta{S_{1}^{(1)}}\geq \Delta{S_{1}^{(2)}}\geq \cdots$.
Let $^{(r)}S_1= S_1-\Delta{S_{1}^{(1)}}- \cdots- \Delta{S_{1}^{(r)}}$, $r=0,1,2,\ldots$,  be the ``trimmed" subordinator.

Define a random point measure on the Borel sets of $(0,1)$ by
\be \label{defB}
\BB^{(r)} = \sum_{i \ge 1} \delta_{J_i(r)}, \quad \text{where} \quad
J_i(r) = \frac{\Delta S_1^{(r+ i)}}{\Delta S_1^{(r)}},\ i = 1, 2, \ldots, \ r\in\N.
\ee
Thus   $\BB^{(r)}$ is defined in terms of {\it ratios of successive jumps} of $^{(r)}S_1$, rather than in terms of the jumps themselves. 
In Proposition 2.1 of \cite{trimfrenzy} it is proved  that 
$\BB^{(r)}$ as defined is distributed as a negative binomial point process with a 
measure  whose density is $\alpha x^{-\alpha-1}\rmd x{\bf 1}_{\{0<x\le 1\}}$.
Note that, by comparison with \eqref{LM2}, this density is truncated to $(0,1]$. 
\eqref{defB}  motivates the further definition: 
%$\BN(r, \Lambda^*)$, where $\Lambda^*(\rmd x)=\alpha x^{-\alpha-1}\rmd x {\bf 1}_{0<x< 1}$. 
\begin{definition}\label{deff}
For each $r=1,2,\ldots$, the  vector 
\bea \label{sub_alpha}
%%V^{(r)} := 
\big(V_n^{(r)}\big)_{n\ge 1} :=
% \big( V_1^{(r)},\,  V_2^{(r)}, \ldots \big)=  
 \Big( \frac{\Delta S_1^{(r+1)}}{{}^{(r)}S_1},\, \frac{\Delta S_1^{(r+2)}}{{}^{(r)}S_1}, \ldots \Big)
=
 \Big( \frac{J_r(1)}{\sum_{i\ge 1}J_i(r)},\, \frac{J_r(2)}{\sum_{i\ge 1}J_i(r)}, \ldots \Big)
\eea
on $\nabla_\infty$ 
is said to have a $\PD_\alpha^{(r)}$ distribution. 
%%When $r =0$, $\PD_\alpha$ is recovered.
\end{definition}
%Thus by \eqref{sub_alpha} we can write vectors in $\PD^{(r)}_\alpha$ in the form
%\[
%\big(V_n^{(r)}\big)_{n\ge 1} 
%= \big( V_1^{(r)},\,  V_2^{(r)}, \ldots \big)=  \Big( \frac{J_r(1)}{\sum_{i\ge 1}J_i(r)},\, \frac{J_r(2)}{\sum_{i\ge 1}J_i(r)}, \ldots \Big).
%\]
%It is easy to see that $\PD_\alpha^{(r)}$ is equivalent to the $\PK^{(r)}(\Lambda^*)$ that arises naturally from such a probabilistic structure.
Although the derivation of $\PD_\alpha^{(r)}$ via the normalised jumps of a subordinator requires $r$ to be a positive integer, the formulae in what follows are valid for any $r>0$ with the obvious interpretations,  and subsequent results hold in this generality. 
%The extra parameter $r$ adds flexibility for modelling purposes,

We refer to \cite{trimfrenzy, IpsenMaller2018} for further details on the stick-breaking properties, Palm characterisation and joint densities of the size biased permutations
of distributions constructed from negative binomial processes.

\section{Ewens Sampling Formula for a $\PD_\alpha^{(r)}$-Partition}\label{ss4}
%%The Ewens sampling formulae for a $\PD^{(r)}_\alpha$-partition is needed for the data analysis in Section \ref{data}.
Using the notation in Subsection \ref{ss3}, consider a sample of size $n$ from a population with infinitely many  blocks, whose proportions are governed by the $\PD_\alpha^{(r)}$ distribution.
Corresponding to a given partition $\Pi_n=(A_1,\ldots,A_k)$ of $n$ into $k$ blocks, 
the components  $M_j$ of the vector $\bfM$ represent the number of blocks (allele types)
of $\Pi_n$ in the sample which have exactly $j$ representatives.
%, and collect the $M_j$ into a vector $\bfM = (M_1,\ldots,M_n)$.
%%is the {\it blocks count vector}.  The
%%random variables $ M_1,\ldots,M_n$ are non-negative, integer-valued and satisfy  $\sum_{j=1}^{n}M_j=k$ and  $\sum_{j=1}^{n}j M_j = n$.
%Let $n_i$ be the number of representatives of block type $i$, $1\le i\le k$, in the sample, with $\sum_{i=1}^k n_i=n$.

%Given  a vector $\bfm=(m_1,m_2,\ldots,m_n)$
%of  non-negative integers with $\sum_{j=1}^{n} m_j=k$ and 
%$\sum_{j=1}^{n}j m_j=n$,  let

 %Ewens' sampling formula %%%(\cite{Antoniak:1969},)(\cite{Ewens1972})
%gives the probability of any particular realization of the blocks count vector sampled from the $\PD(0,\theta)$ distribution.
%We can specialise Theorem \ref{EPPF1} the EPPF for  $\PD_\alpha^{(r)}$ and use it obtain the Ewen's sampling formulae for $\PD_\alpha^{(r)}$. 
The next theorem presents  a version of  Ewens' sampling formula for $\PD_\alpha^{(r)}$.
%, in a form convenient for computations.
%Denote  the ascending factorial by $r^{[k]} := r(r+1)\cdots(r+k-1)$, $k\in\N$.

\begin{theorem}\label{th:4.1}
The Ewens random partition structure derived from $\PD_\alpha^{(r)}$ for a sample of size $n$ partitioned into $k$ blocks has distribution 
\bea\label{4.1}
&&\PP(\bfM=\bfm:=(m_1,m_2,\ldots,m_n))\cr
&&= \alpha^k r^{[k]}
\prod_{j=1}^{n}\frac{1}{m_j!}\Big(\frac{\Gamma(j-\alpha)}{j!}\Big)^{m_j}
 \times
\int_{0}^{\infty}
\frac{\lambda^{\alpha k-1}}{\Psi(\lambda)^{r+k}}
\prod_{j=1}^{n} \big(G_{j-\alpha}(\lambda)\big)^{m_j} \rmd \lambda\,
{\bf 1}_{\{\sum_{j=1}^{n} m_j=k\}},\cr
&&
\eea
where $\sum_{j=1}^{n}j m_j=n$,  $r^{[k]}= r(r+1)\cdots(r+k-1)$
denotes the ascending factorial, 
\be\label{defPsi2}
\Psi(\lambda):=1+\alpha\int_{0}^{1}(1-e^{-\lambda x}) x^{-\alpha-1}\rmd x,
\ee
and
 $G_{j-\alpha}(\lambda)$ is the incomplete gamma distribution function defined by
\be\label{Gdef}
G_{j-\alpha}(\lambda):=\frac{1}{\Gamma({j-\alpha})}\int_{0}^{\lambda}x^{{j-\alpha}-1}e^{-x}\rmd x,\  \lambda>0.
\ee
% and  $\Psi$ is defined in \eqref{defPsi2}. %,  $\bar{\Psi}(\lambda)=\Psi(\lambda)-1$, 
%and $\bar{\Psi}^{(n_i)}(\lambda)=\rmd^{n_i}(1-\bar \Psi(\lambda))/\rmd \lambda^{n_i}$.
\end{theorem}

\medskip\noindent{\bf Remarks.}
(i)\ 
A proof of Theorem \ref{th:4.1} is in Appendix A.1.
Formula \eqref{4.1} looks forbidding (and is computationally challenging) but in fact has a readily interpretable and reasonably tractable structure as we show in Appendix A.2.

(ii)\ 
It's important to note that the integral in \eqref{defPsi2}  is
truncated to $(0, 1)$, whereas integrals in the original $\PD_\alpha$ class are over the whole range $(0, \infty)$.   This is a significant distinguishing feature between the two classes.

\section{Parameter Estimation Methodology}\label{PEM}
To estimate the  parameters in  $\PD_\alpha^{(r)}$  we
minimise the weighted sum of squares
\be\label{WSS}
S:= \sum_{j=1}^n w_j\big(m_j-E_j)^2
\ee
for variations in $\alpha$ and $r$. Here $w_j>0$ are weights (we chose $w_j=j$) and $E_j=E_j(\alpha,r,k)$ is the expected
 value of the allele type count $M_j$, conditional on $\sum_{j=1}^nM_j=k$, based on the current values of $\alpha$ and $r$ and the   $\PD_\alpha^{(r)}$ model.
We have no analytical formula for the $E_j$, so  we  estimate them by simulation. 

This is done   as follows. 
Given starting values $\alpha$ in $(0,1)$, $ r>0$ and a number of blocks, $k$, sample from $\PD_\alpha^{(r)}$ by generating
realisations of $(\Gamma_j)_{j\ge 1}$ random variables (as successive sums of i.i.d. unit exponential rvs), then setting
\be\label{tilpsam0}
\wt p_j= \frac{ \Gamma_{j}^{-1/\alpha}}
{ \sum_{\ell=1}^J \Gamma_\ell^{-1/\alpha}}, \ j=1,2,  \ldots, J,
\ee
where $J$ is a large number; we used $J=1000$ and $5000$.
%.\footnote{We took $J=5000$. A similar method was used %%in the case $r=0$
%by \cite{Labadi2014} as an efficient way of generating observations on  $\PD_\alpha$.}
Then proceed as follows. 

\begin{figure}[h!]
    \centering
    \begin{subfigure}[t]{0.4\textwidth}
        \centering
        \includegraphics[scale = 0.3]{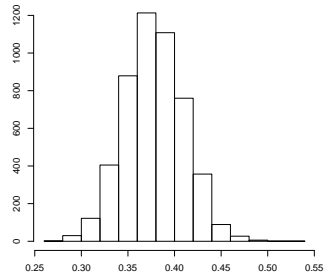}
        \caption{\capfont $\wh\alpha$, Batalling quolls, Locus 3.3.1}
    \end{subfigure}%
    ~ 
    \begin{subfigure}[t]{0.4\textwidth}
        \centering
        \includegraphics[scale = 0.3]{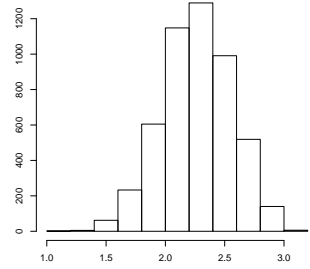}
        \caption{\capfont $\wh r$ Batalling  quolls, Locus 3.3.1}
    \end{subfigure}
    ~
    \begin{subfigure}[t]{0.4\textwidth}
        \centering
        \includegraphics[scale = 0.3]{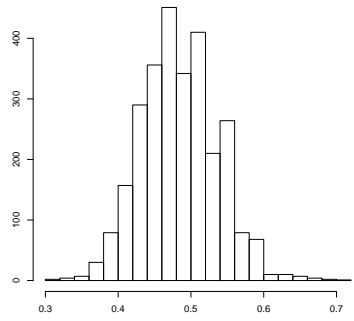}
        \caption{\capfont $\wh\alpha$,  Perth quolls, Locus 3.3.1}
    \end{subfigure}%
    ~ 
    \begin{subfigure}[t]{0.4\textwidth}
        \centering
        \includegraphics[scale = 0.3]{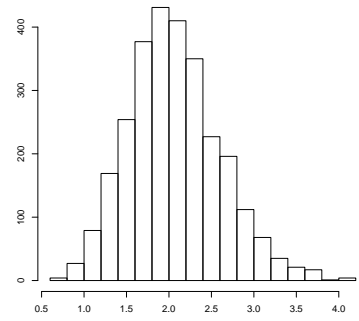}
        \caption{\capfont $\wh r$  Perth quolls, Locus 3.3.1}
    \end{subfigure}

    \caption{Resampling Histograms of Estimates $(\wh  \alpha, \wh  r)$  for Quolls, Locus 3.3.1 }\label{FigQ3.3.1}
\end{figure}

\begin{enumerate}

\item For  the given number of blocks, $k$,   set population probabilities $(p_1,\ldots,p_k)$ as
\be\label{psam0}
p_j= \frac{\wt p_{r+j}}
{ \sum_{\ell=1}^k \wt p_{r+\ell}}, \ j=1,2,  \ldots,k,
\ee
based on the $\wt p_j$ in \eqref{tilpsam0}.

\item   Follow the procedure outlined in %Section \ref{rp} 
Appendix A.1
to form a partition by drawing $k$ observations on a random variable $Y$ having an arbitrary diffuse base distribution $G_0$ which  fixes the block labels
(we used $G_0=N(0,1)$).  %We can think of $G_0$. 
For a sample of size $n$, select $n$ members $(X_i)_{1\le i\le n}$ with replacement from those $k$ numbers according to the probabilities $(p_1,\ldots, p_k)$.
Use the $X_i$ to partition $n$ into integers $n_1,\ldots, n_k$, where $n_i=\#\{\ell: X_\ell=X_i\}$,
$1\le i\le k$. 
Then set $\what m_j=\#\{i:n_i=j,\, 1\le i\le k\}$, $1\le j\le n$.
These form the first replacement sample, $(\what m_j(\ell),\, 1\le j\le n)$, indexed  as $\ell=1$.

\item  Repeat this step $L$ times (we used $L=1000$), obtaining the estimates $\what m_j(\ell)$, for $1\le j\le n$, $\ell=1,2,\ldots, L$.
We describe this procedure as ``resampling".   After this calculate 
\be\label{Efind}
\wh E_j= \frac{1}{L} \sum_{\ell=1}^{L} \what m_j(\ell), \ j=1,2, \ldots, n.
\ee

\item During Steps 1 and 2, we can obtain resampling estimates of the variability of  $\alpha$ and $ r$
by estimating them at each stage  to obtain $(\hat\alpha(\ell), \hat r(\ell))$, $\ell=1, \ldots, L$.
The standard errors in Tables  \ref{Q1} and  \ref{Q2}, % and \ref{tab:4},
  and the histograms in Figure \ref{FigQ3.3.1} %and \ref{Fig1-4} 
were calculated from these samples.

\end{enumerate}

The $\wh E_j$ in \eqref{Efind}, calculated with the  current values of $\alpha$ and $r$,  are substituted for  the expected values in \eqref{WSS}.   
Minimising the quantity  $S$ in \eqref{WSS} is done by a grid search over
 values $\alpha=0.05, 0.1, \ldots, 0.95$ and $r=0.1,0.2,\ldots$.
%The minimum is usually found close to the first stage pair $(\hat\alpha_S, \hat r_S)$  and a finer search is done in this region to focus in on  the estimates.
%which thus form good starting estimates for the procedure.
The resulting values $\what\alpha_n$ and $\what r_n$ are our weighted least squares estimates. Note that this analysis is conditional on the observed number $k$ of species. 

The method is very highly computer intensive but has the virtue of providing as close a fit as possible to the data,  in a weighted least squares sense.
%Standard errors of the estimates are calculated as described in Step 4.\footnote{Strictly speaking these are the estimated standard errors of  $(\what\alpha_S,\what r_S)$ rather than of the overall least squares estimates $(\wh\alpha, \wh r)$; but they are a reasonable approximation for the latter.}
%In the next section we report on the analyses of three rather disparate sets of data,  obtaining estimates of $\alpha$ and $r$ for each, with
%estimates of their precision obtained by the resampling method.
In Appendix A.2 we present an argument to suggest that the estimates $(\wh \alpha_n,\wh  r_n)$ 
will be approximately normally distributed in large samples. 
This is consistent with what we observe empirically in the resampling histograms. 
%% The properties of the estimates thus obtained can be investigated by simulations.

\section{Quolls Data Analysis}
%
%\smallskip\noindent{\it 1. Quolls Data.}\
The paper by Firestone et al. (2000) %\cite{Firestone2000} 
contains genetic data on 4 species of quolls (tiger, eastern, northern and western  quolls)  sampled in 20 locations across Australia. 
At each location allele counts at 6 genetic loci are available, with around 15-20 allele types per locus. 
We concentrate on the western quoll data which are from 2 locations in Western Australia 
(Perth and Batalling State Forest) and consider loci numbers 1.3 and 3.3.1.
 There were a maximum of 15  and 34 individuals available in the 2 locations.
 We chose these particular locations for analysis  from the available data as they have the most allelic  variability, and contain the larger numbers of animals.
  We refer to Firestone et al. (2000) for further information on the methods of collection of samples and other interesting aspects of the data. 

%\begin{table}[h]
%\footnotesize
%\setlength{\tabcolsep}{3pt}   
%\begin{center}
%\begin{tabular}{c|c|cccccccccccccccccccc}
%\hline
%Perth&$j$ &1&2&3&4&9  \\
%&$m_j$ &2&2&1&3&1   \\
%\hline
%Batalling&$j$ &1&3&5&6&9&13&15&16   \\
%&$m_j$ &1&1&1&1&1&1&1&1    \\
%\hline
%\end{tabular}
%\caption{\capfont Quolls data: Locus No. 1.3}\label{Q0}
%\end{center}
%\end{table}
%
%\begin{table}[h]
%\footnotesize
%\setlength{\tabcolsep}{3pt}   
%\begin{center}
%\begin{tabular}{c|c|cccccccccccccccccccc}
%\hline
%Perth&$j$ &1&2&3&5&6 \\
%&$m_j$ &7&1&1&2&1   \\
%\hline
%Batalling&$j$ &1&2&3&7&10&11   \\
%&$m_j$ &4&1&2&1&3&1    \\
%\hline
%\end{tabular}
%\caption{\capfont Quolls data: Locus No. 3.3.1}\label{quoll.3.3.1}
%\end{center}
%\end{table}

% 
% \begin{table}[h]
%\footnotesize
%\centering
%\setlength\tabcolsep{5pt}
%\begin{minipage}{0.48\textwidth}
%\centering
%\begin{tabular}{c|ccccc}\hline
%\hline
%$j$ &1&2&3&4&9  \\
%$m_j$&2&2 &1&3&1 \\
%\hline
%\end{tabular}
%\caption{\capfont Gene Locus 1.3 Perth }
%\label{tab:5}
%\end{minipage}
%\hfill
%\begin{minipage}{0.48\textwidth}
%\centering
%%%\tablewidth=\textwidth
%\begin{tabular}{c|cccccccc}\hline\hline
%$j$ &1&3&5 &6&9 &13&15 &16\\
%$m_j$ &1&1&1&1&1&1&1&1 \\
%\hline
%\end{tabular}
%\caption{\capfont Gene Locus 1.3(BSF)}
%\label{tab:6}
%\end{minipage}%
%\end{table}

 \begin{table}[h]
\footnotesize
\centering
\setlength{\tabcolsep}{3pt}   
%\begin{center}
\begin{minipage}{0.48\textwidth}
\centering
\begin{tabular}{c|c|ccccccccc}
\hline
Perth&$j$ &1&2&3&4&9 &&&& \vline\\
&$m_j$ &2&2&1&3&1  &&&& \vline \\
\hline
Batalling&$j$ &1&3&5&6&9&13&15&16  & \vline \\
&$m_j$ &1&1&1&1&1&1&1&1 & \vline  \\
\hline
\end{tabular}
\caption{\capfont Quolls, Locus 1.3}\label{Q0}
\end{minipage}
\hfill
\begin{minipage}{0.48\textwidth}
\centering
%%\tablewidth=\textwidth
\begin{tabular}{c|c|cccccccc}
\hline
Perth&$j$ &1&2&3&5&6  && \vline \\
&$m_j$ &7&1&1&2&1  && \vline  \\
\hline
Batalling&$j$ &1&2&3&7&10&11 & \vline   \\
&$m_j$ &4&1&2&1&3&1  & \vline   \\
\hline
\end{tabular}
\caption{\capfont Quolls, Locus 3.3.1}\label{tab:6}
\end{minipage}%
\end{table}

In Table \ref{Q0}, for Locus  1.3, we see in the first column for Perth, 
that there were 2 allele types having 1 representative, 
2 allele  types  having 2 representatives, $\ldots$,
1 allele type  having 9 representatives.
  Thus the more common alleles occur in the right of the tables, rarer alleles are in the left.
(The tables omit values of $j$ where $m_j=0$.) 
Total numbers of alleles at Locus 1.3   for Perth were  $n=30$ with  $k=9$  distinct allele types.
Similarly,  for Batalling,  $n=68$ and $k=8$, with the numbers in Table \ref{Q0} for Locus  1.3.
Table \ref{tab:6} has the information for Locus  3.3.1, both locations.
 Since the animals are diploids, the numbers $n$ are double the numbers of individual animals sampled.

Tables \ref{Q1} and \ref{Q2} contain the parameter estimates and their standard errors obtained by the method explained in Section \ref{PEM}. 

\begin{table}[h]
\footnotesize
\setlength{\tabcolsep}{3pt}   
\begin{center}
\begin{tabular}{c|c|cccc}
\hline 
&\# Resamples&$\what{\alpha}$ &\SE($\wh\alpha$) &$\what{r}$&\SE($\wh r$) \\\hline\hline
Perth&$N=1000$ &0.4158 &0.0410&1.16 &0.19\\
&$N=5000$ &0.4159&0.0390&1.15&0.19\\
\hline
Batalling&$N=1000$&0.2990 &0.0208&2.25 &0.20\\
&$N=5000$ &0.2980&0.0206&2.26&0.20\\
\hline
\end{tabular}
\caption{\capfont Estimates and  standard errors,  Locus 1.3}\label{Q1}
\end{center}
\end{table}
\begin{table}[h]
\footnotesize
\setlength{\tabcolsep}{3pt}   
\begin{center}
\begin{tabular}{c|c|cccc}
\hline 
&\# Resamples&$\what{\alpha}$ &\SE($\wh\alpha$) &$\what{r}$&\SE($\wh r$) \\\hline\hline
Perth&$N=1000$ &0.4810& 0.0510 &2.07& 0.52\\
&$N=5000$ &0.4814& 0.0512& 2.08& 0.53\\
\hline
Batalling&$N=1000$&0.3770 &0.0320&2.28&0.29\\
&$N=5000$ &0.3780&0.0310&2.27&0.29\\
\hline
\end{tabular}
\caption{\capfont Estimates and  standard errors,  Locus 3.3.1}\label{Q2}
\end{center}
\end{table}
Based on these estimates we see that,
% $r$ is significantly positive for both locations and both loci, and 
at Locus 1.3, $r$ is significantly larger for Batalling than for Perth, while $\alpha$ is significantly larger for Perth than for Batalling. At Locus 3.3.1, the $r$ estimates are not significantly different between Perth and Batalling, but 
for $\alpha$, the difference approaches significance (larger for Perth).
Thus  %, based on this very scant data, 
there is some evidence that Locus 1.3 discriminates between Perth and Batalling, whereas Locus 3.3.1 discriminates marginally at best.  %; and there is not a great  difference in diversity between locations.
We discuss the interpretation of the parameters further in Subsection \ref{inter}. 

Figure \ref{FigQ3.3.1} displays the resampling histograms (5000 resamples) for the Locus 3.3.1 estimates. Even given such a small data set, the distributions are reasonably close to normal and their spreads are well summarised by the standard deviations in Table \ref{Q2}. 
For Locus 1.3, we omit the histograms which were essentially similar but less well determined as is to be expected from the small sample size at this locus.

\subsection{Interpretation of parameters $\alpha$ and $r$}\label{inter}
%%\noindent {\it Negative Binomial vs Poisson.}\
The important distinguishing feature of the $\PD_\alpha^{(r)}$ distribution over $\PD_\alpha$ is its allowance for over-dispersion in the data. This arises from its being based on an underlying negative binomial point process rather than a  Poisson point process.
%, and can be thought of as a generalisation  of $\PD_\alpha$ in a similar way as the negative binomial distribution allows for ``overdispersion" (variance greater than mean), compared with the Poisson distribution, in the analysis of ordinary count data.
%We begin this discussion with  the beetles data, since the numbers are much larger than for the quolls and the estimates are more reliable.
To investigate   what features of  the data 
the parameter estimates capture, we compare the $\alpha$  values with those  obtained from a fit of the original Kingman-Perman $\PD_\alpha$ model.    The parameter $\alpha$ in $\PD_\alpha$ can be regarded as a measure of  diversity
(``$\alpha$-diversity" \cite{Pitman2006}, p.71).
%Turning to the quolls data, 

For Locus 1.3,  there is significant  evidence of differences between locations  in the estimates of both $\alpha$ and
$r$ from  $\PD_\alpha^{(r)}$  (Table \ref{Q1}).
 At this locus, for  $\PD_\alpha^{(r)}$ we have 
$\wh \alpha=0.4158$ (SE=0.0410)  for Perth, and $\wh\alpha=0.2190$ (SE=0.0208) for Batalling.
For comparison, the  $\alpha$ estimates for the $\PD_\alpha$ model for this locus are:
\ben
{\rm for\ Perth:}\  \wh\alpha=0.3880\ ({\rm SE}=0.1208);
\quad   {\rm for\ Batalling:}\ \wh\alpha=0.2676\ ({\rm SE}=0.0990).
\een 
These  are  quite similar to those from the $\PD_\alpha^{(r)}$ model, but the standard errors for $\wh \alpha$ from $\PD_\alpha$ are considerably larger than from  $\PD_\alpha^{(r)}$. 
 Interpreted as a diversity measure, the $\alpha$ estimates from either model suggest  that  diversity is greater for Perth  than for  Batalling for this locus.
To demonstrate this alternatively, we can  calculate the Shannon entropy index for each location, using the formula
  \be\label{divin}
^{1}H =
-\sum_{i=1}^{k} \pi_i\ln \pi_i
\ee
where $\pi_i=n_i/n,$ $1\leq{i}\leq{k}$, are the allele proportions.\footnote{See \cite{Sherwinetal2017}
for a recent review of the use of entropy and related indices in measuring allelic diversity.}
For  Perth we obtain $1.99$ for $^{1}H$, for  Batalling we get $1.86$.
Although not significantly different these are in accord with the $\wh\alpha$ values.

%Information Theory Broadens the Spectrum of Molecular Ecology and Evolution
%W.B. Sherwin,1,2,* A. Chao,3 L. Jost,4 and P.E. Smouse5
%Trends in Ecology & Evolution, December 2017, Vol. 32, No. 12 https://doi.org/10.1016/j.tree.2017.09.012

%%Most data is available for   
Now consider gene Locus  3.3.1.
The  $\alpha$ estimates for the $\PD_\alpha$ model for this are:
\ben
{\rm for\ Perth:}\  \wh\alpha=0.5260\ ({\rm SE}=0.1134);
\quad   {\rm for\ Batalling:}\ \wh\alpha=0.4026\ ({\rm SE}=0.1009).
\een 
These compare with estimates $0.4814$  (SE=0.0512) and $0.3718$    (SE=0.0310), respectively, for  $\PD_\alpha^{(r)}$
(Table \ref{Q2}).
%%With such a small sample we are reluctant to make definite statements, but similar to the beetles data 
We see that the 
$\wh \alpha$   for  $\PD_\alpha$ do not  differ significantly between locations, and 
 are  quite similar to those from the $\PD_\alpha^{(r)}$ model.
Again the standard errors for $\wh \alpha$ from $\PD_\alpha$ are considerably larger than from  $\PD_\alpha^{(r)}$. 
Likewise,  the estimates of $r$ from  $\PD_\alpha^{(r)}$,
namely, $\wh r=2.08$ (SE=0.52)  for Perth, and $\wh r=2.27$ (SE=0.29) for Batalling, 
 do not  differ significantly between locations,  for Locus 3.3.1.

%%\bigskip {\bf Estimates of $\alpha$  for Perman $\PD_\alpha$ model:}\
%The corresponding $\alpha$ estimates for the beetles data  are:
%\ben
%{\rm for\ 1967:}\  \wh\alpha=0.2290\ ({\rm SE}=0.0490);
%\quad   {\rm for\ 1968:}\ \wh\alpha=0.2240\ ({\rm SE}=0.0520).
%\een 
%%(and of course there is no value for $r$).

%%There are two points to note: (i)\ the values are not  significantly different between years, so  
%(ii)\  the $\alpha$ estimates for the $\PD_\alpha$ model
%% are  quite similar to those from $\PD_\alpha^{(r)}$ but the standard errors from $\PD_\alpha$ are larger by a factor of 10 compared to those from  $\PD_\alpha^{(r)}$. 
%on the other hand, the estimates of $r$ from  $\PD_\alpha^{(r)}$ 
%are significantly different between locations,  providing discrimination.

For the parameter $r$ we can provide a nice intuitive interpretation. 
From Gregoire's formulation  of the negative binomial process we see that, for  
%disjoint Borel subsets $B_1,B_2$ of $\mathbb{R}^{+}$,
%The distributions of the numbers $\mathbb{B}^{(r)}(B_1)$ and $\mathbb{B}^{(r)}(B_2)$ have the properties 
a Borel subset $A$ of $(0,\infty)$, 
\be\label{exp}
\EE(\mathbb{B}^{(r)}(A))=r \Lambda(A),
\ee
%\be\label{cov}
%{\rm Cov}(\mathbb{B}^{(r)}(B_1),\mathbb{B}^{(r)}(B_2))
%=
%r\big(\Lambda(B_1\cap B_2)+\Lambda(B_1)\Lambda(B_2)\big),
%\ee
where $\Lambda(\rmd x)=\alpha x^{-\alpha-1}\rmd x{\bf 1}_{0<x < 1}$, and 
%From these we see that the 
\be\label{var}
{\rm Var}(\mathbb{B}^{(r)}(A))=r\big(\Lambda(A)+\Lambda^{2}(A))
\ee
(see  Proposition 3.3 of \cite{Gregoire1984}).
So expectations and variances  are proportional to $r$, and 
%and so is the covariance in \eqref{cov}.
in \eqref{var} we also see the  extra-Poisson dispersion effect of the negative binomial.
%%Although the mechanism may not be transparent, t
This scaling effect of $r$ transfers through to the data analyses.

 $\PD_\alpha^{(r)}$ does not reduce to  $\PD_\alpha$ when $r=0$; rather, the convergence of 
 $\PD_\alpha^{(r)}$ to a Poisson-Dirichlet model takes place as $r\to\infty$  
(see \cite{IpsenShemehsavarMaller2018}), consistent with the convergence of a negative binomial variable to a Poisson for large values of the appropriate index. 
There is no evidence of the $r$ values going out of bounds in Tables \ref{Q1} and \ref{Q2}.

In summary,  there is reasonable evidence that both $\alpha$ and $r$ in the
 $\PD_\alpha^{(r)}$ model discriminate between locations for Locus 1.3,  but not for  Locus 3.3.1. 
But  the $\PD_\alpha$ model provides no discrimination at either locus. 
By contrast, Firestone et al. (2000) found a tendency for, but no significant difference in, allelic diversity between the Perth and Batalling quolls.

\section{Simulations}\label{sims}

In this section we report on some simulations aimed at
examining the properties of  the  parameter estimates.
%% found in the data analyses of Section \ref{data}.
The procedure is the same as described in  Section \ref{PEM} but we stipulated 
 ``true" values $\alpha=0.2$ and $r=1$ rather than estimated ones.
%we sampled from the  class  $\PD_\alpha^{(r)}$ by generating
%realisations of a $\Gamma_j$ random variable (as successive sums of i.i.d. unit exponential rvs), $j=1,2,\ldots$, then setting
%\be\label{tilpsam}
%\wt p_j= \frac{ \Gamma_{j}^{-1/\alpha}}
%{ \sum_{\ell=1}^J \Gamma_\ell^{-1/\alpha}}, \ j=1,2,  \ldots, J
%\ee
%where $J$ is a large number.
%A similar method was used %%in the case $r=0$
%by Al Labadi and Zarepour (2016) as an efficient way of generating observations on  $\PD_\alpha$.
%
%%
%using ratios of jumps of a trimmed stable subordinator, as follows.
%A  subordinator of index $\alpha\in (0,1)$ was simulated using
%
%its jumps up till time $1$ were ranked in decreasing order, then
%the largest $r$ jumps  were deleted from the subordinator.
%The remaining jumps, taken as ratios of the trimmed  subordinator,
%form a population vector $(p_1,p_2,\ldots)$.We took $\alpha=0.2$ and  $r=1$ as the true values. 
There are two levels of variability to consider, which we  term ``population" and ``resampling" variability.

First we describe the resampling scheme, then how the population simulation is superimposed on it.
The ``resampling" part is as described in Steps 1--4 of Section \ref{PEM}: having fixed a number of blocks, $k$,  set the population probabilities $(p_1,\ldots,p_k)$ as in \eqref{psam0}, based on the $\wt p_j$ in \eqref{tilpsam0}.
Draw the $L$ samples $(\wh m_j(\ell)),\, j=1,2,\ldots, n)$,
$\ell=1,2,\ldots, L$, and from them find least squares estimates,
 denoting them as $(\wh\alpha_1, \wh r_1)$.  These form the first simulation estimates, indexed  as $q=1$.
Repeat this entire procedure $Q$ times to  obtain the simulation sample  $(\wh\alpha_q, \wh r_q)$,
$q=1,2,\ldots, Q$.   We can average these replications  to form a simulation estimate of the original $(\alpha,r)=(0.2,1)$ and display them in a histogram to assess precision.
%The results in Table \ref{tab:10} were summarised from these simulations.

At  the ``population" simulation stage, repeat this entire procedure.   %% $M$ times.  
Still with the same ``true" $\alpha$ and $r$,  draw a new set of $\wt p_j$
values from \eqref{tilpsam0}, recalculate the corresponding $p_j$ from \eqref{psam0}, and then resample $Q$ times as described in the resampling stage. Repeat this whole procedure a total of $N$ times.
%The results in Table \ref{tab:10} were summarised from these simulations.

The resulting entries  in Table \ref{tab:11} are the averages over the $N=100$ population simulations of the $L=1000$  resampling estimates in each population simulation.

\begin{table}[h] 
\footnotesize\center
\begingroup
 \setlength{\tabcolsep}{3pt}
\begin{tabular}{c|c c|c c|c c}
    \hline
 %    \# Redraws $N$, \# Blocks $k$
    Resampling &\multicolumn{2}{|c|}{ $k=2$}& \multicolumn{2}
        {|c|}{$k=5$} &\multicolumn{2}{|c }{ $k=10$}\\ 
 estimates & $n=20$&  $n=40$&$n=100$&$n=200$&$n=200$& $n=400$\\ \hline\hline
   $\wh {\alpha}$ &0.198&0.185&0.221&0.210&0.254&0.239\\ \hline
    $\SE_R{(\wh {\alpha})}$&0.023&0.032&0.037&0.030&0.075&0.055\\ \hline
    $\SE_P(\wh {\alpha})$&0.028&0.033&0.052&0.043&0.086&0.059\\ \hline
$\wh {r}$ &1.24&1.23&1.25&1.31&1.19&1.43\\ \hline
    $\SE_R(\wh {r})$&0.45&0.52&0.47&0.56&0.32&0.66\\ \hline
    $\SE_P(\wh {r})$&0.49&0.53&0.51&0.58&0.36&0.67\\ \hline
\end{tabular}
\endgroup
\caption{\setlength{\baselineskip}{0.8\baselineskip}\small Simulation results: True values: $\alpha=0.2$, $r=1$. $1000$ resamples (R) and $100$ population redraws (P). } \label{tab:11}
\end{table}

%
%
%\begin{table}[h!]     \label{tab:10}
%    \begin{tabular}{|c|c|c|c|c|c|c|}
%    \hline
%     Sample Size, $n$, \\
%\# Blocks, $k$&\multicolumn{2}{|c|}{ n=20, k=2}& \multicolumn{2}
%        {|c|}{n=50, k=5} &\multicolumn{2}{|c|}{n=100, k=10}\\ \cline{1-7}
%       
%  \# Resamples  & N=1000&  N=5000&  N=1000& N=5000&  N=1000& N=5000\\ \hline
%   $\wh {\alpha}$ &0.20024&0.19362&0.26701&0.272466&0.33035&0.32127\\ \hline
%    $SD_R{(\wh {\alpha})}$&0.03448&0.03681&0.07525&0.10231&0.13483&0.13263\\ \hline
%    $SD_P{\wh {\alpha}}$&0.038642&0.03241&0.094149&0.08172&0.14275&0.12512\\ \hline
%$\wh {r}$ &1.34733&1.42204&1.43629&1.3237&1.56653&1.67991\\ \hline
%    $SD_R{\wh {r}}$&0.56431&0.62779&0.53120&0.52164&0.73459&0.86129\\ \hline
%    $SD_P{\wh {r}}$&0.60102&0.59641&0.59946&0.44429&0.77763&0.82139\\ \hline
%    \end{tabular}
%    \caption{\it Simulation Study: Table entries are estimates and their standard errors obtained from 1000 or 5000 resamples and 100 population redraws.  Resampling (R), Population (P).}
%\end{table}

\section{Discussion}\label{Dis}

\smallskip\noindent {\it Properties of the $\PD_\alpha^{(r)}$  estimators.}\
 Given the paucity of the data from which we are trying to make inferences, the estimates are surprisingly good. 
The resampling procedure produces  distributions consistent with the asymptotic analysis in Appendix A.2
which suggests approximate normality should apply in large samples. 
The simulations reinforce this view.
%though their  extent is restricted due to limitations of computational power. 
Overall the results suggest that  the weighted least squares method provides  reliable and robust estimates.

%Regarding the interpretation of the parameter $\alpha$, 
It is notable that the $\alpha$ estimates are very similar for  $\PD_\alpha$ and  $\PD_\alpha^{(r)}$ for the quolls data.
This suggests the presence of an underlying Stable$(\alpha)$ structure for the population in which overdispersion is captured by the introduction of parameter $r$.
We observe that introducing $r$ accounts for much of the variability in estimating $\alpha$.

\smallskip\noindent {\it Other Applications.}\ Besides species and gene sampling models. there are many other applications of Poisson-Dirichlet distributions, among which we mention the areas of Bayesian nonparametrics and  machine learning, which are currently under vigorous development.
We expect the $\PD^{(r)}_\alpha$ model to be useful  as a prior distribution when over-dispersion is present in the data. Many machine learning models such as the latent Dirichlet allocation models (\cite{blei2003}) or mixed membership models (\cite{ghahramani2006}) deal with data, e.g. corpora of documents, which possess large underlying variation. With the flexibility provided by the over-dispersion parameter $r$, one could in principle select a more appropriate prior distribution for analysis.

%Bayesian, machine learning
%$\PD^{(r)}_\alpha$ is a special case of random discrete distributions on the infinite simplex constructed from negative binomial point processes, which as a general class, includes the two parameter Poisson-Dirichlet distributions for $\theta > 0$ (\cite{IpsenMaller2018}). 

\smallskip
\noindent {\it Concluding Remarks.}\ 
 The methods we illustrate potentially have implications for the management and conservation of threatened populations. 
Measuring the level of genetic diversity of a population, and comparing levels between populations, are of major importance in determining conservation strategies 
%of threatened species 
and providing guidance for reintroduction programs. 
The study of Firestone et al. (2000) is the first to attempt to determine the level of diversity of the western quolls, with a specific objective of being able to distinguish genetically between the two subgroups and between other Australian quoll species.
The data in this case is necessarily small-sample, but the  $\PD_\alpha^{(r)}$ analysis contributes useful information regarding diversity even in this case.
%In other applications, we have seen that the model 
%% and, as shown  for the tropical insects data, 
%can provide some quite definite conclusions in  larger data sets.
Similar small-scale data sets are commonly found in the conservation literature: see, e.g., \cite{CPM2006}, \cite{Ketal2018}.
{\it  Without sound knowledge of the genetic diversity within
and between .. % the remaining Tasmanian 
populations ...
%of eastern quolls, 
it is  difficult to determine which populations are valuable sources for reintroductions.}  (Firestone et al. (2000)).

% Campos, J.L.,  Posada, D. and Mor\`an, P. (2006)
%Genetic variation at MHC, mitochondrial and microsatellite loci in isolated populations of Brown 
%trout ({\it Salmo  trutta}).
%Conservation Genetics 7, 515--530.
%
% Khosravi, R., Hemami, M-R.,  Malekian, M., Silva, T.L.,  Rezaei, H.R. and  Brito, R.C. (2018) 
%Effect of landscape features on genetic structure of the goitered gazelle (Gazella subgutturosa) 
%in Central Iran. 
%Conservation Genetics 19, 323--336.

%Stuart B. Piertney, S.B. and Webster, L.M.I. (2010) 
%Characterising functionally important and ecologically meaningful genetic diversity using 
%a candidate gene approach. 
%Genetica,  138,  419--432.

Our  $\PD_\alpha^{(r)}$ model is an extension of the  $\PD_\alpha$ model, which is a subclass of the general  $\PD(\alpha,\theta)$ models due to  \cite{PY1997}.
 Statistical inference for the range of  $\PD(\alpha,\theta)$ models,
even for the one-parameter  $\PD_\alpha$ and $\PD(\theta)$ models, is still 
little understood. \cite{Carlton1999} gives some discussion of estimation problems in these particular models. 
Our  investigations suggest that the two-parameter  $\PD_\alpha^{(r)}$  holds  promise of being a practical and useful alternative.

\renewcommand{\bibfont}{\small}
\bibliography{Library_Levy_Apr2018.bib}
\bibliographystyle{newapa}

%Blei, D. M., Ng, A. Y., & Jordan, M. I. (2003). Latent dirichlet allocation. Journal of machine Learning research, 3 (Jan), 993–1022.
%Ferguson, T. S. (1973). A Bayesian analysis of some nonparametric problems. The Annals of Statistics, 209–230.
%Ghahramani, Z. & Griffiths, T. L. (2006). Infinite latent feature models and the Indian buffet process. In Advances in Neural Information Processing systems, (pp. 475–482).

%Hjort, N. L. (1990). Nonparametric Bayes estimators based on beta processes in models for life history data. The Annals of Statistics, 18 (3), 1259–1294.

%Patil, G. & Taillie, C. (1982). Diversity as a concept and its measurement. Journal of the American Statistical Association, 77 (379), 548–561.

\section{Appendix A.1: The $\PD_\alpha^{(r)}$ Methodology}\label{A.1}
%%\subsection{Species Sampling Models and Random Partitions}\label{rp}
Here we outline the technology leading to \eqref{4.1}, referring to the basic theory developed in \cite{trimfrenzy, IpsenMaller2018} where necessary.
%We need some background in  the theory of random partitions.
Let $(P_i)_{i\in\N}$ be a random sequence in $\nabla_\infty$ and 
$G$ a random discrete distribution generated from it.  %%$(P_i)_{i\in\N}$.
With  $\delta_x$ denoting a point mass at $x\in\R$, 
$G$  can be written as
\be\label{Fdef}
G =\sum_{i\ge 1}P_i \delta_{Y_i}, 
\ee
where $P_1\ge P_2 \ge \cdots$, and the $(Y_i)$  are independent and identically distributed (i.i.d.) random variables,
 independent of the $(P_i)$,  having a ``base" distribution $G_0$.

Let $(X_i)$ be independent samples from $G$. They generate a random partition $\Pi$ consisting of non-overlapping ``blocks", by assigning $i$ and $j$ to the same block when $ X_i=X_j$. 
Then, conditionally given $G$, the $X_i$  are i.i.d. with distribution $G$. 
We call $G$ a {\it species sampling model}.
%%; see for example \cite{ishwaran2003} for further discussion of these ideas.

%Formally, the random partition $\Pi$ is identified with the sequence $(\Pi_n)$, where
Let  $\Pi_n$ be the restriction of $\Pi$ to the finite set $\N_n:=\{1,2,\ldots,n\}$. The distribution of  $\Pi_n$ is such that for each partition $\{A_1,A_2,\ldots,A_k\}$ of  $\N_n$ with $\# A_i=n_i$, $1\leq{i}\leq{k}$, we have
\be \label{2.1}
\PP(\Pi_n=\{A_1,A_2,\ldots,A_k\})=:p(n_1,\ldots, n_k)
\ee
for some symmetric function $p(\cdot)$ of sequences of positive integers,
where $n_i\geq{1}$ and $\sum_{i=1}^{k}n_i=n$. The function $p(n_1,\ldots, n_k)$ is called the {\it exchangeable partition probability function} (EPPF) of $\Pi$.

With $P_i>0$ denoting the size of the $i$th largest atom of a species sampling model $G$, as in \eqref{Fdef}, 
let 
$\wt P_i$ denote the size of the $i$th atom discovered in the process of random sampling; equivalently, $\wt P_i$ is the asymptotic frequency of the $i$th class of $\Pi$ in order of its appearance in the sample.
The sequence $(\wt P_i)$ is then a {\it size biased permutation} of $(P_i)$, and
the EPPF in \eqref{2.1} satisfies (\cite{Pitman2003}, Eq. (3))
\be \label{2.2}
p(n_1,\ldots, n_k)
=\mathbb{E}\Bigg(\prod_{i=1}^k\wt P_i^{n_i-1}\prod_{i=1}^{k-1}\Big(1-\sum_{j=1}^{i}\wt P_j\Big)\Bigg).
\ee
%We refer to \cite{James2002, James2005} 
% for some explicit EPPF formulae  for general completely random measures obtained through Poisson partition calculus.
%A recent paper by \cite{RDG2010} gives a model for species abundance, allowing for spatial distribution, based on the model \eqref{Fdef} and the associated stick-breaking process. 

%%\subsection{$\PK^{(r)}$--Partitions and size biased sampling}\label{PSBS}

Now let $\{A_1, \ldots, A_k\}$ be a partition of $\NN_n$ generated by $\PD_\alpha^{(r)}$
% from a negative binomial process $\BN(r, \Lambda)$ with representation $\BB^{(r)}$ as in \eqref{pointform},  for some $r > 0$ and $\Lambda$ satisfying \eqref{con1}. 
 and let $(\wt J_i)_{i\in\N}$ be the  size biased permutation of the sequence $\JJJ:=(J_i(r))_{i\in\N}$ defined in \eqref{defB}; 
thus, conditionally on $\JJJ$,
$\wt J_1$ takes value  $J_i(r)$ with probability $J_i(r)/{}^{(r)}T$;
while for $n\ge 1$,
conditional on $\JJJ$ and $\{ \wt J_1, \ldots, \wt J_n \}$, $\wt J_{n+1}$ takes value  $J_j(r) \in \JJJ\setminus\{ \wt J_1, \ldots, \wt J_n \}$ with probability
$J_j(r)/({}^{(r)}T- \sum_{i=1}^n \wt J_i \big)$.
Here  ${}^{(r)}T$ %%:= T(\BB^{(r)})$ be 
is the sum of the points in   $\BB^{(r)}$.
%%remaining after $n$ size-biased picks.
%% from $\BB^{(r)}$ be the sequence $({}^{(r)}T_n)_{n\ge 0}$;
%%thus,  ${}^{(r)}T := {}^{(r)}T_0 := T(\BB^{(r)})$.
%, and,  for each $n\ge 1$,
%\be \label{Tdef}
%{}^{(r)}T_n := \, {}^{(r)}T_{n-1} - \wt J_n\,.
%\ee
%%Then (see \cite{trimfrenzy})  $^{(r)}T$  
It has density $g_{r}(t)$  (see \cite{trimfrenzy} and compare with \eqref{1.1}) 
%%:= \PP\big( T(\BB^{(r)}) \in \rmd t\big)/\rmd t$ 
satisfying
\be \label{gdL}
\int_0^\infty e^{-\lambda x} g_{r}(x) \rmd x =
\Big(1 + \int_{0}^{\infty} (1- e^{-\lambda x}) \Lambda(\rmd x)  \Big)^{-r}, \ \lambda>0.
\ee

The next lemma 
%is a version of Lemma 5 of \cite{Pitman2003} for our setup.   %key for our development. %It
 gives a formula for the joint distribution of the  $\PD_\alpha^{(r)}$-partition with the size biased $\wt J_i$ and their sum.
Recall $\rho(x)= \alpha x^{-\alpha-1}{\bf 1}_{\{0<x\le 1\}}$ in \eqref{LM2}. 
%$ {}^{(r)}T$,  which differs from that of Pitman by inclusion of the parameter $r$ in the formula. 
%This is a nontrivial distinction as we remark at the end of this section.
%Denote  the ascending factorial by $r^{[k]} := r(r+1)\cdots(r+k-1)$, $k\in\N$, and recall  $\N_n=\{1,2,\ldots,n\}$. 
%Lemma \ref{lm2.1} is proved in Appendix A.1. 

\begin{lemma}\label{lm2.1}
%Assume $\Lambda$ has a density $\rho$.
%Let $\Pi_{n}$ be the restriction to $\N_n$ of a  $\PK^{(r)}(\rho)$--partition $\Pi$ 
%based on $(\wt J_i)_{i\in\N}$,  the {\it size biased permutation} of the ordered points  $(J_{(i)})$ of an NBPP, $\BB^{(r)}$.   
%Then, f
For a partition $\{A_1,A_2,\ldots,A_k\}$ of $\N_n$
such that $\# A_i=n_i$, $1\leq{i}\leq{k}$, we have
\bea \label{2.4}
&&
\PP\big(\Pi_n=\{A_1,A_2,\ldots,A_k\},\; \wt J_i \in \rmd x_i,\;1\leq{i}\leq{k},\; {}^{(r)}T\in \rmd t\big)\nonumber\\
&=&
r^{[k]}\; t^{-n}g_{r+k}\Big(t-\sum_{i=1}^{k}x_i\Big)\; \rmd t\;\prod_{i=1}^{k}\rho(x_i)x_{i}^{n_{i}} \rmd x_i,
\eea
where $0\le x_i\le 1$, $x_i\ne x_j$,  $1\le i,j \le k$, and $t>\sum_{i=1}^kx_i$.
\end{lemma}
%Although the derivation of $\PD_\alpha^{(r)}$ via the normalised jumps of a subordinator requires $r$ to be a positive integer, \eqref{gdL} is defined for any $r>0$ and Lemma \ref{lm2.1} and subsequent results hold in this generality.

\noindent {\bf Proof of Lemma \ref{lm2.1}:}\
%	Let $\Pi_n = \{A_1, \ldots, A_k\}$ be a $\PK^{(r)}(\rho)$ partition restricted to $\N_n$ and $(\wt W_i^{(r)})$ the size-biased permutation of $(W_i^{(r)})$ distributed as $\PK^{(r)}$, as in \eqref{1.2}. 
%	Then, a
%	As in \eqref{2.2}, conditionally on $(\wt W_i^{(r)})_{1\le i\le k}$ we have
%	\begin{align}\label{partition1}
%		&\PP\big(\Pi_n =\{A_1, \ldots, A_k\} \,|\, \wt W_1^{(r)}, \ldots, \wt W_k^{(r)} \big) \nonumber \\ 
%	&= \prod_{i=1}^{k-1}  
%	\Big(\big(\wt W_i^{(r)}\big)^{\# A_i-1}\big(1-\sum_{\ell=1}^{i}\wt W_\ell^{(r)}\big)\Big) \big(\wt W_k^{(r)}\big)^{\# A_k-1}.
%	\end{align}
%%%Recall that 
%The joint distribution of $(\wt W_i^{(r)})_{1\le i\le k}$ is determined by the joint distribution of  $(\wt J_1, \ldots, \wt J_k)$ and $ {}^{(r)}T$.
%%, where the $(\wt J_i)$ are a size-biased permutation of the ordered  points $J_{(i)}$ of a point process $\BB^{(r)}$ in $\BN(r, \rho)$ and ${}^{(r)}T = \sum_{i} \wt J_i$.
%Thus \eqref{partition1} implies 
For  $\PD_\alpha^{(r)}$, the $\wt P_i$ in \eqref{2.2},
 conditional on $\wt J_i$ and $^{(r)}T$, 
 are replaced by  $\wt J_i/^{(r)}T$, so we have for the LHS of \eqref{2.4}
\begin{align*}  %%\label{2.4a} 
\prod_{i=1}^{k-1}  \Big[\Big(\frac{x_i}{t}\Big)^{\# A_i-1}\Big(1-\sum_{\ell=1}^{i}\frac{x_\ell}{t}\Big)\Big]
 \Big(\frac{x_k}{t}\Big)^{\# A_k-1}
\times \PP\Big( \wt J_i \in \rmd x_i,\, 1\leq{i}\leq{k},\; {}^{(r)}T\in \rmd t\Big).
\end{align*}
Letting $n_i:=\#A_i$,  $1\le i\le k$, with $\sum_{i=1}^kn_i=n$, the first factor on the RHS  equals
\be\label{A}
\frac{1}{t^{n-1}} \Big( \prod_{i=1}^k x_i^{n_i-1}\Big)
\Big(\prod_{i=1}^{k-1}\big(t-\sum_{\ell=1}^ix_\ell\big)\Big).
\ee
Using Proposition 2.3 in \cite{trimfrenzy}, the second factor on the RHS  equals
\be\label{B}
 r^{[k]}\, g_{r+k}\Big(t-\sum_{i=1}^{k}x_i\Big)\; \rmd t\,
\prod_{i=1}^{k}
\frac{\rho(x_i)x_i \rmd x_i}{t-\sum_{\ell=1}^{i-1}x_\ell}.
\ee
%(recall that $g_k(t)\rmd t= P(^{(k)}T\in\rmd t)$).
Multiply \eqref{A} and \eqref{B} after observing that
\ben
\prod_{i=1}^{k} \frac{1}{t-\sum_{\ell=1}^{i-1}x_\ell}= 
\frac{1}{t} \prod_{i=2}^{k} \frac{1}{t-\sum_{\ell=1}^{i-1}x_\ell}= 
\frac{1}{t} \prod_{i=1}^{k-1} \frac{1}{t-\sum_{\ell=1}^{i}x_\ell}
\een
(with the convention $\sum_1^0=0$) to get \eqref{2.4}.     \halmos

%Alternate proof of Lemma 2.1 in ISM version 28 Feb 2018.
%
%Let $\Pi$ be constructed by laying random intervals $I_i$ down on the interval $[0,{}^{(r)}T]$ in some arbitrary random order, where the lengths $J_{r}(i)=\vert{I_i}\vert$ are the ranked ratio points of the negative binomial process with intensity density $\rho(x) $, and ${}^{(r)}T=\sum_{i}J_{r}(i)$.
%Let $\Pi$ be the partition of $\N$ generated by the random equivalence relation $i\sim j $ iff either $i=j$ or ${}^{(r)}TU_i$ and ${}^{(r)}TU _j$ fall in the same interval $I_i$ for some $i$, where $U_1,U_2,\ldots,$ are  i.i.d Uniform$[0,1]$ random variables.
%
%For the event in the lefthand side of  \eqref{2.4} to occur:
%
%\noindent
%{\rm (i)}
%There must be a negative binomial point in each $\rmd x_i$, for each $1\leq{i}\leq{k}$; and\\
%{\rm (ii)}
%given (i), the sum of the remaining negative binomial points  ${}^{(r)}T$ must fall in an interval of length $\rmd t$ near $ t-\sum_{i=1}^{k}x_i$;
%and \\
%{\rm (iii)}  given (i) and (ii), for each $1\leq{i}\leq{k}$ and each $m\in A_{i}$ the sample point ${}^{(r)}TU_{m}$ must fall in the interval of length $x_i$.
%
%The infinitesimal probability in \eqref{2.4}  therefore equals
%\ben %%\label{2.6}
%r^{[k]}g_{r+k}\Big(t-\sum_{i=1}^{k}x_i\Big)\rmd t
%\prod_{i=1}^k\rho(x_i) \rmd x_i\prod_{i=1}^k\Big(\frac{x_i}{t}\Big)^{n_i}.
%\een
%Since $n_1+n_2+\cdots +n_k=n$, this is the same as the righhand side of \eqref{2.4}.  \halmos

Next we derive  expressions for the EPPF %of a random partition 
generated by 
$\PD_\alpha^{(r)}$. %$\PK^{(r)}(\rho)$.
%The main result, Theorem \ref{EPPF1},  is proved in Appendix A.1. 

%\subsection{EPPF for $\PD_\alpha^{(r)}$}
%The construction in  Lemma \ref{lm2.1} allows for a quite general
%intensity density function $\rho(x)$. In later sections we consider
%special cases  where $\rho(x)$ is proportional to $\alpha x^{-\alpha-1}{\bf 1}_{0<x < 1}$.
%These correspond to the original Poisson-Kingman class $\PD_\alpha$, and to the new $\PD_\alpha^{(r)}$ class, respectively.\footnote{\cite{IpsenMaller2018} also consider $\rho_\gamma(x){\bf 1}_{x>0}$, a density based on a Gamma subordinator. This gives rise to  a class  $\PK^{(r)}(\rho_\gamma)$ related to Kingman's $\PD(\theta,0)$ class. But we do not consider this here.}
%But for the present section we continue with a general $\rho$ on $(0,\infty)$. 
%In the following, 

\begin{theorem}\label{EPPF1}
Two formulae for the  {\rm EPPF} of a  $\PD_\alpha^{(r)}$--partition are:
\be \label{2.5}
p(n_1,\ldots,n_k)=r^{[k]}\int_{x_1 =0}^{\infty}\cdots\int_{x_k =0}^{\infty}
\int_{w=0}^{\infty}\frac{g_{r+k}(w)\prod_{i=1}^{k}\rho(x_i)x_i^{n_i}}{(w+\sum_{i=1}^{k}x_i)^{n}}\rmd w\, \rmd x_1\cdots \rmd x_k
\ee
and
\be\label{3.13}
p(n_1,\ldots,n_k)=\frac{\alpha^k r^{[k]}}{\Gamma{(n)}}
\prod_{i=1}^{k}\Gamma(n_i-\alpha)\times 
\int_0^{\infty}
\frac{\lambda^{k\alpha-1}}{\Psi(\lambda)^{r+k}}
\prod_{i=1}^{k}G_{n_i-\alpha}(\lambda)\, \rmd \lambda,
\ee
where
 $\Psi$ is defined in \eqref{defPsi2} and
 $G_{n_i-\alpha}(\lambda)$ is the incomplete gamma function.
 % in \eqref{Gdef}.
%\be\label{defPsi2}
%\Psi(\lambda)=1+\alpha\int_{0}^{1}(1-e^{-\lambda x}) x^{-\alpha-1}\rmd x
%\ee
%and
% $G_{n_i-\alpha}(\lambda)$ is the incomplete gamma distribution function defined by
%\be\label{Gdef}
%G_{n_i-\alpha}(\lambda):=\frac{1}{\Gamma({n_i-\alpha})}\int_{0}^{\lambda}x^{{n_i-\alpha}-1}e^{-x}\rmd x,\  \lambda>0.
%\ee

%\end{corollary}

\end{theorem}
%
%\medskip\noindent {\bf Proof of Corollary \ref{cor3}:}\
%%Eq. \eqref{3.12} is derived by substituting $\alpha x^{-\alpha-1} {\bf 1}_{\{0<x<1\}}$ for $\rho(x)$ in \eqref{Jdef}, then applying \eqref{gibbs} of  Proposition  \ref{cor2}; 
%\eqref{3.13} follows from \eqref{2.6}
%after substituting $\alpha x^{-\alpha-1} {\bf 1}_{\{0<x<1\}}$ for $\rho(x)$ in \eqref{2.7aa} and \eqref{2.7b},
%The integral in \eqref{2.7b}, restricted to $(0,1)$, can then be written in terms of the incomplete gamma function after a change of variable.
%\halmos

%%\medskip\noindent{\bf Remarks.}
%We note the important distinction: the integrals in \eqref{bark} and \eqref{defPsi2} are truncated to $(0,1)$, whereas those in \eqref{2.7a} and \eqref{2.7aa} are over the whole range $(0,\infty)$.  
%%(i)\ 
%It's important to note that the integral in \eqref{defPsi2}  is
%truncated to $(0, 1)$, whereas integrals in the original $\PD_\alpha$ class are over the whole range $(0, \infty)$.   This is a significant distinguishing feature between the two classes.
%In comparing \eqref{3.13} with Corollary 6 of \cite{Pitman2003}, we note that we cannot take $r=0$ in %\eqref{2.4}, 
%\eqref{3.13} to get that the EPPF for $\PD_\alpha^{(r)}$ reduces to the corresponding  formulae for $\PD_\alpha$;  \eqref{2.4} is based on the {\it ratios} of the jumps in \eqref{defB}, and these are not defined for $r=0$.  Nevertheless, the class of distributions $\PD_\alpha^{(r)}$ {\it does} contain  $\PD_\alpha$  when $r=0$.

 \medskip\noindent {\bf Proof of Theorem \ref{EPPF1}:}\
 Use the change of variable $t-\sum_{i=1}^{k}x_i=w$ in \eqref{2.4}, then
 integrate out with respect to $x_i$, $1\leq{i}\leq {k}$, and $w$, to get \eqref{2.5}. 
Formula \eqref{3.13} follows by applying the identity 
\be\label{ID}
\frac{1}{x^n}=
\frac{1}{\Gamma(n)}\int_{0}^{\infty}\lambda^{n-1}e^{-\lambda x}\rmd \lambda,
\ee
valid for any $x>0$, 
%\eqref{ID}
 to \eqref{2.5} to deduce
\bean
&&
\int_{w=0}^{\infty}\frac{g_{r+k}(w)\prod_{i=1}^{k}\rho(x_i)x_i^{n_i}}{(w+\sum_{i=1}^{k}x_i)^{n}}\rmd w\cr
&&\cr
&&= 
\frac{1}{\Gamma(n)} \int_{w=0}^{\infty}\int_{\lambda=0}^\infty
\lambda^{n-1} e^{-\lambda(w+\sum_{i=1}^{k}x_i)})
g_{r+k}(w)\rmd w\, \rmd \lambda\cr
&&\cr
&&
=
\frac{1}{\Gamma(n)} \int_{\lambda=0}^\infty(\Psi(\lambda))^{-r-k}\prod_{i=1}^{k}e^{-\lambda x_i}\rmd\lambda
\quad  {\rm (by}\ \eqref{gdL}). 
\eean
Multiply this by 
$r^{[k]}  \prod_{i=1}^{k}\rho(x_i)x_i^{n_i}$ then integrate with respect to the $x_i$ to obtain \eqref{3.13}.    \halmos

\medskip\noindent {\bf Proof of Theorem \ref{th:4.1}:}\
Choose  $\bfm=(m_1,m_2,\ldots,m_n)$ such that  $\sum_{j=1}^{n} m_j=k$ and $\sum_{j=1}^{n}j m_j=n$.
%Due to the natural bijection between partitions of $n$ and such  vectors,
%of counts $(m_j)_{1\leq{j}\leq n}$, for a vector of non-negative integers subject to  $\sum_{j=1}^{n} m_j=k$ and $\sum_{j=1}^{n}j m_j=n$, 
The probability that $\Pi_n$ has $m_j$ blocks of size $j$, $1\leq{j}\leq{n}$, is 
\ben %\label{4.1a}
\PP(\bfM=\bfm) =N_k(\bfm) p(n_1,\ldots, n_k)
 =\displaystyle{\frac{ n!}{ \prod_{j=1}^{n} j!^{m_j}m_j!}} p(n_1,\ldots, n_k),
\een
where  $N_k(\bfm)$
%\ben
%N_k(\bfm) =\displaystyle{\frac{ n!}{ \prod_{j=1}^{n} j!^{m_j}m_j!}}
%\een
is the number of orderings of the sample prescribed by $\bfm$.
Substitute for $p(n_1,\ldots, n_k)$  from \eqref{3.13}, then
observe that
$\prod_{i=1}^kG_{n_i-\alpha}(\lambda)=\prod_{j=1}^n\big(G_{j-\alpha}(\lambda)\big)^{m_j}$, 
and adjust the constant factor, to get  \eqref{4.1}.
%Note again that $\Psi(\lambda)$ %and $\bar \Psi(\lambda)$ in
%is the truncated version  in \eqref{defPsi2}. % \eqref{4.1}.
%% is obtained by multiplying that expression with \eqref{4.1a}.
\halmos

\section{Appendix A.2: Asymptotics of Estimators}\label{A.2}
In this section we derive a useful representation for the distribution of the species counts as a mixture of independent Poisson rvs, and use it to suggest
%a proof of 
the asymptotic distribution of the least squares estimators.
Write \eqref{4.1} in the form
\bea\label{4.1aa}
\PP(\bfM=\bfm)  
&=&
 \int_{0}^{\infty}
\frac{ r^{[k]}  \lambda^{k\alpha-1}}{\Psi(\lambda)^{r+k}}
\prod_{j=1}^{n}\frac{1}{m_j!}
\left(\frac{1}{j!}  \int_{0}^{\lambda}\alpha x^{j-\alpha-1}e^{-x}\rmd x\right)^{m_j}
 \rmd \lambda \cr
 &&\cr 
 &=&
 \int_{0}^{\infty} \frac{ r^{[k]}   \lambda^{k\alpha-1}}{\Psi(\lambda)^{r+k}}
\prod_{j=1}^{n}\left(
 \frac{e^{-F_j(\lambda)}F_j(\lambda)^{m_j}}{m_j!}\right)\, e^{ T_n(\lambda)}
 \rmd \lambda,
 \eea
%In \eqref{4.1aa} 
where the $F_j(\lambda)$ and $T_n(\lambda)$ are defined by
\ben
F_j(\lambda): = \frac{1}{j!}  \int_{0}^{\lambda}\alpha x^{j-\alpha-1}e^{-x}\rmd x
\ \  {\rm and}\ \
T_n(\lambda):= \sum_{j=1}^n F_j(\lambda)
= \alpha  \int_{0}^{\lambda}\sum_{j=1}^n  \frac{x^{j-\alpha-1}}{j!}e^{-x}\rmd x.
\een
%Add the $F_j(\lambda)$ over $1\le j\le n$ and let $n\to \infty$ to see that
%and
%\be\label{ass}
%T_n(\lambda):= \sum_{j=1}^n F_j(\lambda)
%= \alpha  \int_{0}^{\lambda}\sum_{j=1}^n  \frac{x^{j-\alpha-1}}{j!}e^{-x}\rmd x.
%\ee
Let $(N_j(\lambda))_{1\le j\le n}$ be $n$ independent Poisson rvs with
 $\EE N_j(\lambda)=F_j(\lambda)$. Then
\be\label{prep}
\PP(\bfM=\bfm)  
=
 \int_{0}^{\infty}\frac{ r^{[k]}   \lambda^{k\alpha-1}}{\Psi(\lambda)^{r+k}}
P\big(N_j(\lambda)=m_j,\, 1\le j\le n\big)\,  e^{ T_n(\lambda)}\, \rmd \lambda 
 \ee
represents the distribution of $\bfM_n$ as a mixture of independent Poisson rvs.
As $n\to\infty$ we have the finite limit
\be\label{ass2}
\lim_{n\to\infty} T_n(\lambda)=
 \alpha  \int_{0}^{\lambda}x^{-\alpha-1}(1-e^{-x})\rmd x
:= T(\lambda).
\ee

To estimate  $\alpha$ and $r$ we minimise the sum of squares  in \eqref{WSS} by setting its derivative equal to 0. So to obtain  $(\wh\alpha_n, \wh r_n)$
we solve the unbiased estimating equation
\be\label{ee}
\sum_{j=1}^n w_j\big(M_j-E_j(\alpha,r)\big) \frac{\partial E_j(\alpha,r)}{\partial (\alpha,r)}=0.
\ee
A Taylor expansion and standard arguments show that  $(\wh\alpha_n, \wh r_n)$ will be asymptotically normally distributed under some  regularity conditions 
if the LHS of \eqref{ee} is asymptotically normal.

From  \eqref{prep}, we can obtain an expression for  linear combinations such as those in \eqref{ee} 
as corresponding  linear combinations of the $N_j(\lambda)$, and it is then plausible that the estimates will be close to normally distributed after appropriate norming and centering.
Working this out in detail is a highly technical exercise; we have to keep in mind the  restrictions $\sum_{j=1}^n m_j=k$ and $\sum_{j=1}^n jm_j=n$ implicit in 
 \eqref{prep}, and the regularity conditions that have to be imposed are not at all transparent, as is often the case with this kind of analysis. 

  We do not attempt to write these conditions out explicitly since they cannot be checked easily in practice anyway, but the distributions of the estimates shown  in the resampling histograms are reasonably close to  normal with a spread which is accurately described by the resampling standard errors.
 In practice, we should check on closeness to normality by using a resampling scheme such as we suggest, or some other bootstrapping or similar method.
 
%
%
%\newpage
%\bigskip\bigskip
% 
%\begin{figure}[h!]
%    \centering
%    \begin{subfigure}[t]{0.4\textwidth}
%        \centering
%        \includegraphics[scale = 0.3]{quoll_alphaGeneLocus1-3(BSF).eps}
%        \caption{\capfont $\wh\alpha$, Batalling  quolls, Locus 1.3}
%    \end{subfigure}%
%    ~ 
%    \begin{subfigure}[t]{0.4\textwidth}
%        \centering
%        \includegraphics[scale = 0.3]{quoll_rGeneLocus1-3(BSF).eps}
%        \caption{\capfont $\wh r$, Batalling  quolls,  Locus 1.3}
%    \end{subfigure}
%    ~
%    \begin{subfigure}[t]{0.4\textwidth}
%        \centering
%        \includegraphics[scale = 0.3]{quoll_alphaGeneLocus1-3(perth).eps}
%        \caption{\capfont $\wh\alpha$,  Perth quolls, Locus 1.3}
%    \end{subfigure}%
%    ~ 
%    \begin{subfigure}[t]{0.4\textwidth}
%        \centering
%        \includegraphics[scale = 0.3]{quoll_rGeneLocus1-3(perth).eps}
%        \caption{\capfont $\wh r$,  Perth quolls, Locus 1.3}
%    \end{subfigure}
%
%
%    \caption{Resampling Histograms of Estimates $(\wh  \alpha, \wh  r)$  for Quolls, Locus 1.3 }\label{FigQ1.3}
%\end{figure}

%%\newpage
\bigskip\bigskip
 
\end{document}